\newif\ifjpsj
\jpsjfalse   

\ifjpsj
 \documentclass[twocolumn,letter]{jpsj3}
\else
 \documentclass[twocolumn,amsmath,amssymb,longbibliography,prb]{revtex4-2}
 \usepackage{hyperref}
\usepackage[dvipdfmx]{graphicx}

\usepackage{bm}
\usepackage{color}
\usepackage{braket}
\usepackage{lipsum}
\usepackage{mathdots}
\usepackage[version=3]{mhchem} 
\usepackage{ulem}
\usepackage{color, soul}

\usepackage{xspace}
\usepackage{url}

\usepackage{xspace}
\usepackage{url}

\newcommand{\CSS}  {\ce{Co3Sn2S2}\xspace}

\begin{document}
\setulcolor{red}

\title{
Chiral gauge field in fully-spin polarized Weyl semimetal with magnetic domain walls
}

\author{Akihiro Ozawa$^1$}\thanks{akihiroozawa@issp.u-tokyo.ac.jp}
\author{Yasufumi Araki$^2$}\thanks{araki.yasufumi@jaea.go.jp}
\author{Kentaro Nomura$^3$}\thanks{nomura.kentaro@mbp.phys.kyushu-u.ac.jp}

\affiliation{$^1$Institute for Solid State Physics, The University of Tokyo, Kashiwa 277-8581, Japan}
\affiliation{$^2$Advanced Science Research Center, Japan Atomic Energy Agency, Tokai, Ibaraki 319-1195, Japan}
\affiliation{$^3$Department of Physics, Kyushu University, Fukuoka 819-0395, Japan}

\begin{abstract}

Modulation of magnetization in magnetic Weyl semimetals leads to the shift of Weyl points in momentum space,
which effectively serves as the chirality-dependent gauge field for the Weyl fermions.
Here, we theoretically study  such a magnetization-induced chiral gauge field,
in a  fully spin-polarized  Weyl ferromagnet \CSS.
From a tight-binding model of \CSS on stacked kagome lattice with magnetism,
we calculate the magnetization-dependent evolution of the Weyl points in momentum space,
resulting in the chiral gauge field.
In the presence of the magnetic domain wall structure,
we evaluate the chiral magnetic field arising from the spatial profile of the chiral gauge field.
We find that
    a magnetic domain wall in \CSS gives rise to a giant chiral magnetic field for the Weyl fermions,
    which reaches the order of a few hundred tesla to induce the Landau quantization.
    Such a giant chiral magnetic field may also influence the novel transport phenomena,
    such as the charge pumping by the domain wall motion, compatible with the spin-motive force.

\end{abstract}

\maketitle

\section{Introduction}

Weyl fermion is the elemental description of massless fermion in relativistic quantum mechanics \cite{weyl1929gravitation}.
The Weyl fermion has the chirality degree of freedom,
which characterizes its spin being either parallel (right-handed) or antiparallel (left-handed) to its momentum.
To both left-handed and right-handed fermions,
the ordinary U(1) gauge field for electromagnetism, appearing in Maxwell's equations,
couples equally.
On the other hand, 
one may also consider a hypothetical U(1) gauge field coupling oppositely to different chiralities,
which is termed as the chiral (axial) gauge field.
Both the ordinary and chiral gauge fields participate in the chiral anomaly of the Weyl fermions \cite{Bardeen1969PhysRev,Landsteiner2016notes}.
The chiral anomaly comes from the topological nature of the Weyl fermions and demands the non-conservation of fermion chirality.
The chiral anomaly was originally motivated by the decay process of neutral pion
in the context of particle physics~\cite{Adler1969PhysRev,Bell1969pcac}.

While the chiral gauge field is introduced as the hypothetical degree of freedom
in relativistic quantum mechanics,
its effect is expected to be emulated in the class of materials called Weyl semimetals~\cite{ilan2020pseudo}.
In Weyl semimetals, 
Weyl fermions appear as low-energy excitations around the Weyl points,
where the valence and conduction bands touch linearly in momentum space \cite{Murakami2007, Wan2011, Burkov2011, Armitage2018}.
In crystalline systems,
the Weyl points hosting fermions with opposite chiralities necessarily appear in pairs \cite{nielsen1981absence,nielsen1983adler}.
If the left- and right-handed Weyl points are shifted oppositely in momentum space by a perturbation,
the shift can be regarded as the chiral gauge field~\cite{Liu2013PRB}.
Several types of perturbations 
are expected to serve as the chiral gauge field in Weyl semimetals,
such as lattice strain \cite{Cortijo2015,Pikulin2016PRX}, chemical substitution \cite{Kariyado2019jpsj}, and irradiation of circularly polarized light \cite{Ebihara2016,Bucciantini2017}. 
In contrast to the ordinary electromagnetic fields,
the temporal and spatial modulations of the chiral gauge field
serve as chirality-dependent electromagnetic fields for the Weyl fermions,
named as \textit{chiral electromagnetic fields}
(see Appendix \ref{sec:chiral-gauge} for their fundamental descriptions).

In magnetic Weyl semimetals, where time-reversal symmetry is broken by magnetism~\cite{Wan2011,Burkov2011},
    the chiral gauge field may be achieved by the perturbation of magnetic ordering.
    This picture has been justified in the simple model of Weyl fermions,
    which includes the isotropic spin-momentum locking structure and the exchange coupling to the magnetization \cite{Liu2013PRB,Araki2020adp,ilan2020pseudo}.
    In this model, the low-energy Hamiltonian for the Weyl fermions reads,
    \begin{align}
        H_\eta({\bm k}) &= \eta v_{\rm F} {\bm \sigma}\cdot{\bm k} +J {\bm \sigma}
        \cdot {\bm m} 
        = \eta v_{\rm F} {\bm \sigma} \cdot ({\bm k}-\eta e {\bm A}_5),
        \label{eq:minimal}
    \end{align}
    where $\eta = \pm$ is the chirality of the Weyl Fermion, $v_{\rm F}$ is the Fermi velocity, ${\bm \sigma}$ is the Pauli matrix for spins,
    and $J$ is the exchange coupling constant to the magnetization ${\bm m}$.
    Here, it is obvious that the perturbation of magnetization is equivalent to the chiral gauge field ${\bm A}_5=- (J/ev_{\rm F}) {\bm m}$.
    With this spin-momentum locking picture,
    novel magnetoelectric phenomena have been theoretically proposed,
    which interconnect the transport of Weyl fermions with real-space textures and time-dependent dynamics of magnetization~ 
    \cite{ilan2020pseudo,Araki2020adp,kurebayashi2021jpsj,Heidari2023,Harada2023prb}.
    In contrast,
 in the realistic Weyl semimetal materials~\cite{Fang2003,Kuroda2017,Sakai2018,Liu2018,ma2019spin},  
     it is a challenging problem to observe and analyze whether such phenomena are present.
This is because the presence or absence of spin-momentum locking depends on 
the structure of spin-orbit coupling (SOC) and the magnitude of the spin polarization in each material.

    In this work,
    we focus on the ferromagnetic Weyl semimetal material \CSS~\cite{Liu2018,Wang2018,Liu2019},
    and theoretically study the relation between the magnetization and the chiral gauge field therein.
    The Weyl fermion picture works well in \CSS,
    because it exhibits the Weyl points close to the Fermi level with relatively small Fermi surfaces. 
    However, for these Weyl fermions, the spin-momentum locking picture is not applicable.
    This is because the spin splitting by the spontaneous magnetization is dominant over SOC,
    so that the Weyl fermions near the Fermi level are almost spin polarized.
    Therefore, here we study the chiral gauge field structure explicitly in \CSS,
    by calculating its electronic structure based on the effective tight-binding model~\cite{Ozawa2019,Ozawa2022,Lau2023, Seki2023prr, Ozawa2024pra}.
    Since the presence of magnetic textures, especially domain walls, are experimentally reported in \CSS~\cite{Sugawara2019,Lee2022,yoshikawa2022non,wang2023magnetism},
    our idea of incorporating the chiral gauge field picture would help us predict the magnetoelectric responses possible therein.

Our findings in this work are as follows.
We introduce the tight-binding model in Sec.~\ref{sec:tight-binding}.
Then we find that the positions of the Weyl points smoothly change depending on the magnetization direction,
which yields the relation between the chiral gauge field and the magnetization direction~(Sec.~\ref{sec:weyl-trajectory}).
From the obtained relation, 
we show that a magnetic domain wall gives a chiral magnetic field for the Weyl fermions,
whose magnitude can reach up to the order of $260\;\mathrm{T}$ for the domain wall of a width 10nm~(Sec.~\ref{sec:domainwall}).
With such a giant chiral magnetic field,
the sliding motion of the domain wall generates a Hall voltage in its sliding direction
(Sec.~\ref{sec:discussion}).
The induced voltage scales about 800 times larger
than that from the spinmotive force in conventional ferromagnets \cite{Volovik1987jopc,Barnes2007prl}.
Such a magnetoelectric response enables the efficient detection of the domain wall motion,
which would be helpful in designing spintronics devices based on \CSS.

\section{Effective Tight-binding Model}
\label{sec:tight-binding}

\begin{figure}[t]
    \centering
    \includegraphics[width=1.0\hsize]{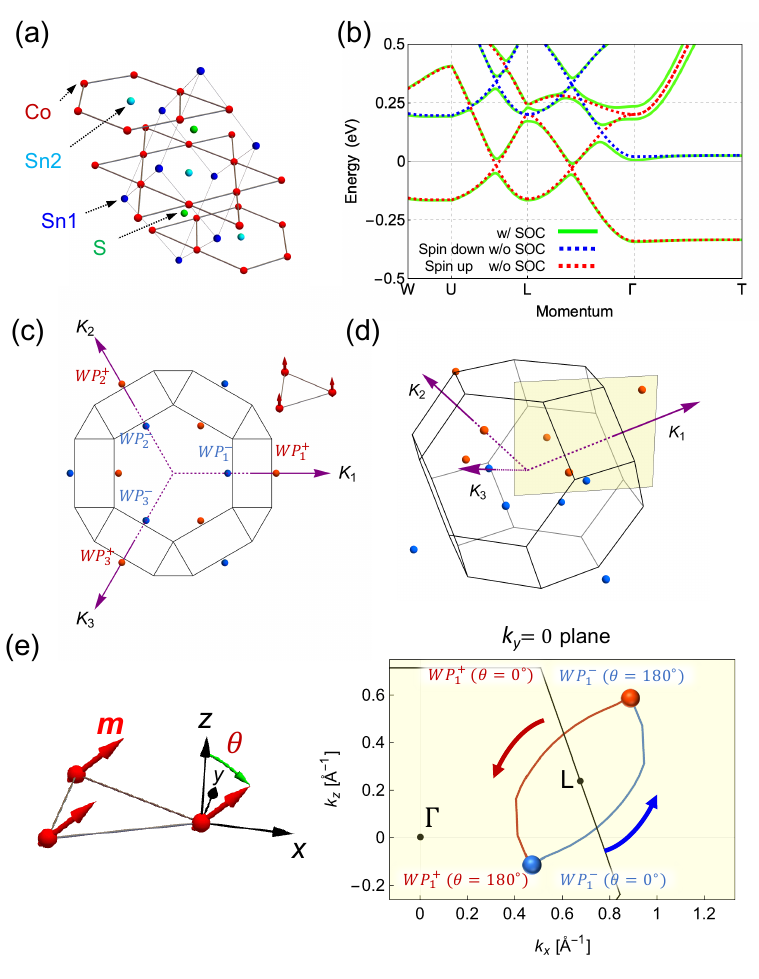}
    \caption{
    (Color online)~
    (a)~The schematic figure of the crystal structure of \CSS.
    (b)~The electronic band structure along the high symmetry line obtained by the effective model~\cite{Ozawa2019}, with perpendicular magnetization $\theta = 0^\circ$.
    The Red~(blue) bands are spin-up ~(spin-down) bands without spin-orbit coupling.
    The green bands are those without spin-orbit coupling.
    (c,d)~The configuration of the Weyl points in the Brillouin zone. 
    Red and blue points distinguish the chirality of the Weyl points.
    (e) The trajectories of the Weyl points $\mathrm{WP}_1^+$ (red) and $\mathrm{WP}_1^-$ (blue),
    with the magnetization direction $\theta$ varied from $0^\circ$ to $180^\circ$ and $\phi = 0^\circ$ fixed.
    Both the two Weyl points move on the $k_y =0$ plane.
    }
    \label{fig:fig1}
\end{figure}

In this section, we introduce the tight-binding model for \CSS,
which we use for the calculations of the Weyl point structure.
The crystal structure of this system is shown in Fig.~\ref{fig:fig1}(a).
Co atoms form kagome layers and are responsible for ferromagnetic ordering.
Sn atoms are located at the center of the hexagons of the kagome lattice.
Two types of triangular lattice layers formed by Sn and S are sandwiched by the kagome layers.
Here we use the tight-binding model of \CSS constructed in our previous works \cite{Ozawa2019,Ozawa2022,Lau2023, Seki2023prr, Ozawa2024pra}.
To describe the low-energy band structure including the Weyl points,
we incorporate the partially occupied atomic orbitals around the Fermi level in the model.
We use one of the $d$ orbitals from each Co atom on the kagome layers and one of the $p$ orbitals from each Sn atom on the triangular layers, which we call Sn1.
    This assumption is valid as long as the crystal field splitting energy is sufficiently large.
    As a consequence,
this model consists of the $(3+1)$ sublattices in each rhombohedral unit cell~[see Fig.~\ref{fig:fig1}(a)].
    The primitive translational vectors are defined as 
    ${\bm a}_1=(\frac{a}{\sqrt{3}},0,\frac{c}{3})$,
    ${\bm a}_2=(-\frac{a}{2\sqrt{3}},\frac{a}{2},\frac{c}{3})$,
    ${\bm a}_3=(-\frac{a}{2\sqrt{3}},-\frac{a}{2},\frac{c}{3})$.
Here, $a$ and $c$ are the lattice parameters of the conventional unit cell, given as 
$a = 5.37$~\AA~and $c = 13.18$~\AA.

The total Hamiltonian of this model is composed of three terms,
\begin{align}
    H_{0}=H_{\text{d-p}}+H_{\rm{so}}+H_{\rm exc}.
    \label{eq:hamiltonian}
\end{align}
Here, 
$H_{\text{d-p}}$ is the hopping term,
$H_{\rm so}$ is the SOC term,
and 
$H_{\rm exc}$ is the exchange coupling term between itinerant electron spins and the mean field.
In the following, we show the structure of each term.

The spin-independent hopping term $H_{\text{d-p}}$ is given by,

\begin{align}
 H_{\text{d-p}}=- \sum_{ijs}t_{ij} d^{\dagger}_{is}d_{js}+t^{\rm dp}\sum_{\langle ij \rangle s}(d^{\dagger}_{is}p_{js}+p^{\dagger}_{is}d_{js})\\ \nonumber
+\epsilon_{\rm{p}}\sum_{is} p^{\dagger}_{is}p_{is}. \nonumber
\end{align} 
Here $d_{is}$ and $p_{is}$ are the annihilation operators of Co-$d$ and Sn1-$p$ orbitals, respectively,
where $i$ and $s$ are the indices for site and spin $(\uparrow,\downarrow)$, respectively.
The first term represents the hoppings between Co sites residing on the kagome layers.
For the hopping amplitudes $t_{ij}$,
we incorporate the hoppings between the first- and second-nearest neighboring sites within each kagome layer,
which we denote as $t_1$ and $t_2$, respectively,
and the interlayer hopping $t_z$  between Co sites.
The second term corresponds to the hybridization $t^{\rm dp}$ between the neighboring Co-$d$ and Sn1-$p$ orbitals.
The third term accounts for the difference of on-site energies between Co-$d$ orbital and Sn1-$p$ orbital.
We fix the on-site energy of Co-$d$ orbital to zero,
and take that for Sn1-$p$ orbital as $\epsilon_{\mathrm{p}}$.

The SOC term is composed of two parts, 
$H_{\rm so}=H^{\rm intra}_{\rm so} +H^{\rm inter}_{\rm so}$,
which are defined as
\begin{align} 
    H^{\rm intra}_{\rm so} &=-i t^{\rm intra}_{\rm so} \sum_{\langle\!\langle ij \rangle\!\rangle ss'} \nu_{ij}   d^{\dagger}_{is} { \sigma}^z_{ss'} d_{js'}, 
    \label{eqn:KM} \\
    H^{\rm inter}_{\rm so} &= -i t^{\rm inter}_{\rm so} \sum_{\langle ij \rangle ss'} {\bm \lambda}_{ij} \cdot d^{\dagger}_{is} {\bm \sigma}_{ss'} d_{js'}.
    \label{eqn:stgRashba}
\end{align}
Here, $\boldsymbol{\sigma}=(\sigma^x,\sigma^y,\sigma^z)$ is the Pauli matrix for the electron spin.
These SOC terms are introduced as the spin-dependent imaginary hoppings,
whose signs depend on the hopping directions.

The first part $H^{\rm intra}_{\rm so}$
is associated with the hoppings between next-nearest neighboring sites within each kagome layer.
It comes from the in-plane local electric field,
whose source is the Sn atom residing on the center of each hexagon in the kagome lattice, which we call Sn2.
The factor $\nu_{ij}=\pm 1$ depends on whether the hopping direction is counterclockwise or clockwise
with respect to the adjacent Sn2 atom.
This term breaks spin SU(2) symmetry,
while it preserves U(1) symmetry by the out-of-plane spin component $\sigma_z$.
Its structure is similar to the SOC term in Kane--Mele model defined on the honeycomb lattice \cite{Kane2005,Guo2009}.

The second part $H^{\rm inter}_{\rm so}$ in $H_{\rm so}$ describes the SOC associated with the inter-layer hopping \cite{Ozawa2022}.
The summation by $\langle ij \rangle $ is taken for neighboring Co sites between adjacent kagome layers.
The vector $\bm{\lambda}_{ij}$ denotes the direction of the effective magnetic field,
which acts on the electron hopping from site $j$ to site $i$.
We define
${\bm \lambda}_{CA}=\frac{{\bm a}_1}{2} \times \frac{{\bm a}_3}{2}/|\frac{{\bm a}_1}{2} \times \frac{{\bm a}_3}{2}|$.
${\bm \lambda}_{AB}=\frac{{\bm a}_2}{2} \times \frac{{\bm a}_1}{2}/|\frac{{\bm a}_2}{2} \times \frac{{\bm a}_1}{2}|$.
${\bm \lambda}_{BC}=\frac{{\bm a}_3}{2} \times \frac{{\bm a}_2}{2}/|\frac{{\bm a}_3}{2} \times \frac{{\bm a}_2}{2}|$.
This term is required by the crystalline structure of this system,
which violates the conservation of U(1) symmetry by spin $\sigma^z$.

The exchange coupling between the spins of itinerant electrons and the mean field is given by,
\begin{align}
    H_{\rm{exc}}=-J \sum_{iss'} {\bm m}_{i} \cdot (d^{\dagger}_{is} {\bm \sigma}_{ss'} d_{is'}
    + p^{\dagger}_{is} {\bm \sigma}_{ss'} p_{is'}).
    \label{eq:exc}
\end{align}
Here $J$ is the exchange coupling constant,
and ${\bm m}_i$ is mean field on site $i$ that is induced by the Coulomb interaction.
    To make the calculation process simple,
    we introduce the same coupling for the Co sites and the Sn1 sites.
    The ground state structure of $\boldsymbol{m}_i$ was determined by the self-consistent calculation,
    which revealed the stable ferromagnetic ordering in the small carrier doping regime~\cite{Ozawa2022}.
    Here we assume the ferromagnetically ordered state~$\boldsymbol{m}_i = \boldsymbol{m}$,
    with $|\boldsymbol{m}|=1$.

In the following calculations, we set $t_1=0.15\;\mathrm{eV}$ as a unit of energy, and define
$t_2=0.6t_1$, 
$t^{\rm{dp}}=1.8t_1$, 
$t_{z}=-1.0t_1$, 
$\epsilon_{\rm p}=-7.2t_1$,
$t_{\rm so}^{\rm intra}=-0.1t_1$,
$t_{\rm so}^{\rm inter}=0.1t_1$,
and $J=1.2 t_1$.
These parameters are chosen to fit the low-energy band structure to that obtained by the previous first-principles calculations \cite{Liu2018, Wang2018,Ghimire2019}.

\section{Magnetization direction dependence of Weyl point structure}
\label{sec:weyl-trajectory}

Based on the effective model constructed in the previous section,
we investigate how the positions of the Weyl points in momentum space depend on the direction of the magnetization.
By calculating the band structure under the uniform magnetization,
we first study the structure of the Weyl points.
    We parametrize the direction of magnetization
    as ${\bm m}=m(\sin{\theta}\sin{\phi},\sin{\theta}\cos{\phi},\cos{\theta})$.
    Here $\theta$ is the polar angle with respect to the $z$-axis,
    and $\phi$ is the azimuthal angle measured from the $x$-axis.
We then visualize the trajectories of the Weyl points, as a function of $\theta$.
We find that, except for some specific $\phi$, the Weyl points move smoothly in momentum space, in accordance with changes in the magnetization direction.

\begin{figure*}[t]
    \centering
    \includegraphics[width=1.0\hsize]{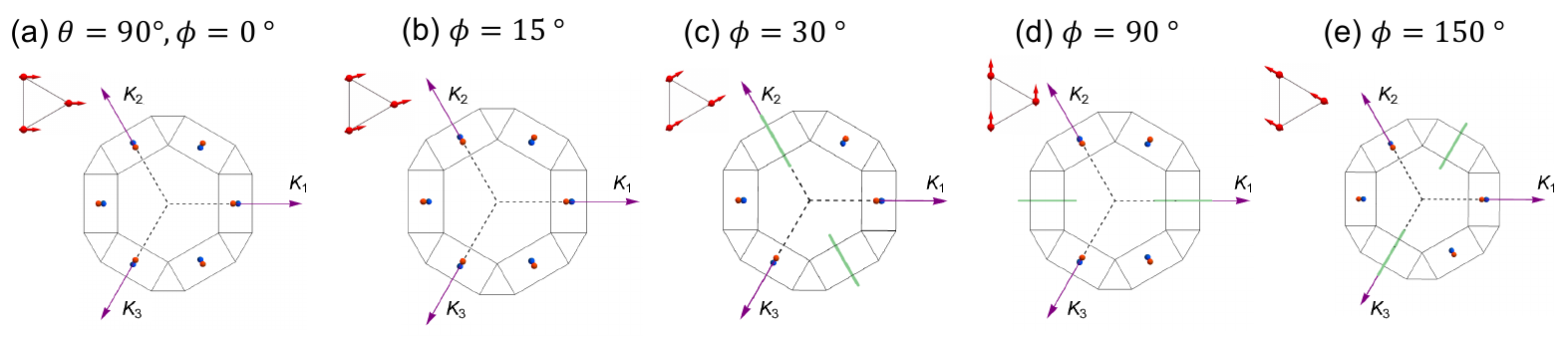}
    \caption{
    (Color online)~
    The in-plane magnetization angle $\phi$-dependences of the nodal structure with tilting angle $\theta=90^\circ$ for different angles, 
    (a) $\phi=0^\circ$
    (b) $\phi=15^\circ$
    (c) $\phi=30^\circ$
    (d) $\phi=90^\circ$
    (e) $\phi=150^\circ$.
    The inset for each is the real-space magnetization structure on the kagome lattice.
    Red and blue points distinguish the chirality of the Weyl points.
    The green lines in (c), (d), and (e) are the gapless nodal rings.
    }
    \label{fig:fig2}
\end{figure*}

    Let us first review the electronic structure of this model in the most typical situation,
    where the magnetization points in the easy axis of the magnetic anisotropy, i.e., $\boldsymbol{m} \parallel \hat{z} \ (\theta = 0^\circ)$.
    The results herein were reported in detail in Ref.~\cite{Ozawa2019}.
By diagonalizing the tight-binding Hamiltonian, 
we obtain the energy eigenvalues $E_n({\bm k})$,
where $n$ is the band index that is labeled in order from 
the bottom $n=1$ to the top $n=8$.
The band structure along the high-symmetry lines is shown in Fig.~\ref{fig:fig1}(b).
In the absence of the SOC term $H_{\rm{so}}$,
both the spin-up and spin-down states,
which are shown by the red and blue lines, respectively,
exhibit the band crossing structure forming nodal rings around three L-points.
Note that this structure applies to an arbitrary magnetization direction $\boldsymbol{m}$,
as long as there is no SOC.
Once we introduce the SOC term, as shown by the green lines in Fig.~\ref{fig:fig1}(b),
these nodal rings are gapped out, except for two Weyl points for each nodal ring,
under $\boldsymbol{m} \parallel \hat{\boldsymbol{z}}$.
The system thus shows three pairs of Weyl points in the whole Brillouin zone,
as shown in Fig.~\ref{fig:fig1}(c) and (d).
We refer to these Weyl points as $\mathrm{WP}_1^\pm$, $\mathrm{WP}_2^\pm$, and $\mathrm{WP}_3^\pm$,
where $\pm$ denotes the chirality of each Weyl point. 
The nodal structures shown here with and without SOC are consistent with those obtained by the first-principles calculations \cite{Liu2018,Wang2018,Ghimire2019}. 
The anomalous Hall conductivity was calculated by this model and it is quantitatively consistent with experiment~\cite{Ozawa2019,Lau2023,Ozawa2024pra}.

We now calculate the nodal structure with the magnetization $\boldsymbol{m}$ tilted from $\theta = 0^\circ$.
Let us consider the cases where $\boldsymbol{m}$ 
points in the in-plane direction
, i.e., $\theta = 90^\circ$,
with $\phi$ varied from $0^\circ$ to $180^\circ$.
(For $180^\circ < \phi < 360^\circ$,
the nodal structure can be obtained by applying the time-reversal operation to that for the azimuthal angle $\phi - 180^\circ$.)
We find that the nodal structure varies depending
on $\phi$, as shown in Fig.~\ref{fig:fig2}.
For the angles 
(c)~$\phi=30^{\circ}$, 
(d)~$\phi=90^{\circ}$, and 
(e)~$\phi = 150^{\circ}$,
we find that one of the nodal rings remains gapless,
while the other two are gapped out to show Weyl points.
The remaining nodal ring resides on the plane perpendicular to ${\bm m}$, which is protected by the mirror symmetry.
For example, for $\phi = 90^\circ$~[Fig.~\ref{fig:fig2}~(d)],
there appears a nodal ring instead of the Weyl points ${\rm WP}^{\pm}_1$.
Here, both the lattice structure and the magnetization $\boldsymbol{m} \parallel \hat{\boldsymbol{y}}$ are invariant
under the mirror reflection by the $xz$-plane.
Therefore, regardless of the presence or absence of the SOC term,
the two bands crossing at the mirror plane $(k_y = 0)$ are the eigenstates of the mirror reflection~\cite{Fang2015prb,Yamakage2016jpsj,Chan2016prb}.
The nodal ring structure on this mirror plane is thus unchanged even though the SOC term is introduced.
This discussion also applies to the other two mirror planes,
corresponding to the cases $\phi=30^\circ$ and $\phi=150^\circ$.
For other $\phi$, such as (a) $\phi=0^{\circ}$ and (b) $\phi=15^\circ$ in Fig.~\ref{fig:fig2},
symmetries by these three mirror planes are violated,
and hence all three nodal rings are gapped out to show Weyl points.

Next, we vary the 
polar angle $\theta$ and investigate the positions of the Weyl points.
Here we focus on one pair of Weyl points ${\rm WP}^{\pm}_1$ residing on the $k_y = 0$ plane,
with the azimuthal angle fixed to $\phi = 0^\circ$.
As a result, we find that the Weyl points move continuously as $\theta$ is varied,
whose trajectories are shown in Fig.~\ref{fig:fig1}(e).
In this figure, the red and blue paths correspond to the trajectories of ${\rm WP}^{+}_1$ and ${\rm WP}^{-}_1$, respectively.
At $\theta=180^{\circ}$, the positions of ${\rm WP}^{+}_1$ and ${\rm WP}^{-}_1$ are switched from those at $\theta=0^{\circ}$.
For $\phi$ other than $0^\circ$, the Weyl points move continuously as a function of $\theta$ as well,
except for $\phi = 90^\circ$ or $270^\circ$,
where the Weyl points reduce to the nodal rings as we have seen above.
The other pairs of the Weyl points follow the similar trajectories,
which we do not explicitly show here.

From the calculations shown in this section,
we find that the variation of the magnetization direction $\boldsymbol{m}$
leads to the positional shift of the Weyl points in momentum space,
although the spin-momentum locking is less dominant.
Therefore, we can regard that the perturbation in $\boldsymbol{m}$ induces the chiral gauge field for the Weyl fermions.
Such a correspondence is helpful in understanding the effects of spatially nonuniform magnetic textures on the Weyl fermions, as we shall see below.

\section{Chiral magnetic field at magnetic domain wall}
\label{sec:domainwall}

With the relation between the magnetization direction and the positions of the Weyl points seen in the previous section,
we now evaluate the configuration of the chiral electromagnetic fields residing at magnetic textures.
In particular, we consider the magnetic domain wall structure in \CSS.
Magnetic domains in \CSS are observed 
by the Lorentz transmission electron microscopy (TEM) \cite{Sugawara2019}
and the magneto-optical Kerr effect (MOKE)\cite{Lee2022}.
The magnetotransport measurement during the magnetization switching process
also gives an implication on the nucleation of magnetic domains \cite{Shiogai2022}.
    Moreover, switching of magnetic domains in \CSS films is also enabled by the application of circularly polarized light \cite{yoshikawa2022non}.
Therefore, it is plausible to consider the chiral gauge field emerging from domain walls in \CSS.
    
To consider the chiral gauge field from magnetic texture, 
let us focus on a single pair of Weyl points on the momentum space positions $\boldsymbol{K}_\eta$ (with the chiralities $\eta =\pm$) in general.
If $\boldsymbol{K}_{+}$ and $\boldsymbol{K}_{-}$ depend on the magnetization direction $\boldsymbol{m}$, 
we can regard that the variation of $\boldsymbol{m}$ yields the chiral gauge field,
\begin{align}
    \boldsymbol{A}_5 = \frac{1}{2e} \left( \boldsymbol{K}_+ - \boldsymbol{K}_- \right), \label{eq:A5-K}
\end{align}
for the Weyl fermions (see Appendix \ref{sec:chiral-gauge}).
Therefore, if there is a real-space texture in $\boldsymbol{m}(\boldsymbol{r})$,
it corresponds to a spatially varying chiral gauge field configuration $\boldsymbol{A}_5(\boldsymbol{r})$,
which yields a chiral magnetic field \cite{Araki2016prb,Araki2020adp},
\begin{align}
    \boldsymbol{B}_5(\boldsymbol{r}) = \boldsymbol{\nabla} \times \boldsymbol{A}_5(\boldsymbol{r}). \label{eq:B5-A5}
\end{align}
Note that the idea of the chiral magnetic field is applicable as long as the variation of $\boldsymbol{m}$ is moderate enough
so that $\boldsymbol{m}$ can be considered \textit{locally uniform}.
This condition is satisfied when the length scale $l_m$ of the magnetic texture is much longer than
the length scale $l_f$ for the Weyl fermions,
\begin{align}
    l_m \gg l_f. \label{eq:local-uniform}
\end{align}
The length scale $l_f$ characterizes how much the electron state keeps its coherence:
$l_f$ is identified as the Fermi wavelength $\lambda_F=1/k_F$ in the ballistic transport regime,
the mean-free path $\lambda_{mf} = v_F \tau$ in the diffusive transport regime (with $\tau$ the transport relaxation time),
the magnetic length $l_\phi=(eB)^{-1/2}$ under the exposure of a strong magnetic field $B$, etc.
We shall revisit this condition later in this section.

\begin{figure}[t]
\centering
\includegraphics[width=1.0\hsize]{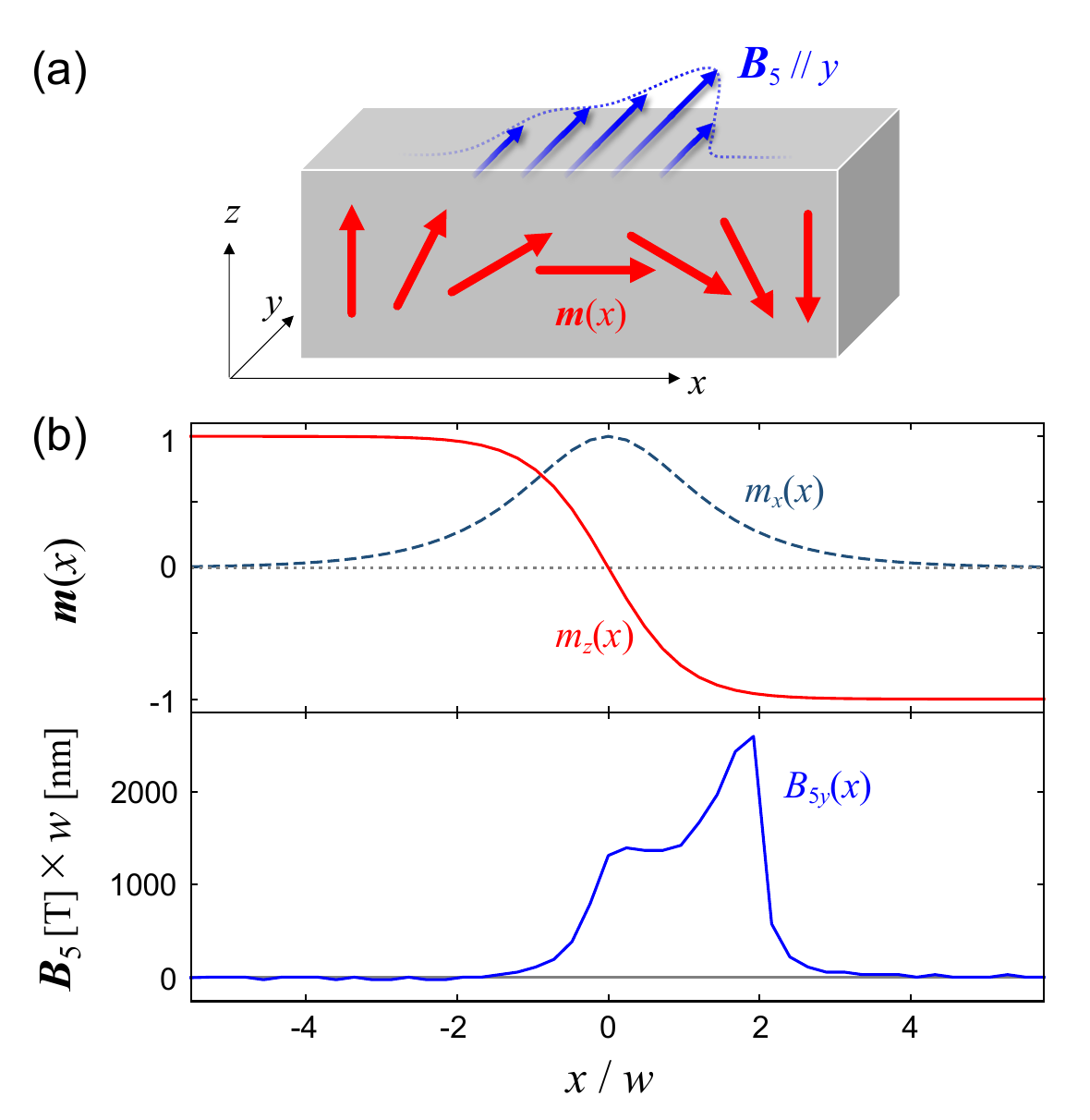}
\caption{(Color online)~
(a) Schematic picture of the spatial configurations of
the N\'{e}el-type magnetic domain wall $\boldsymbol{m}(x)$ and the chiral magnetic field ${\bm B}_5(x)$.
(b) Spatial distribution of $\boldsymbol{m}(x)$ used in the calculations (Top panel),
and the profile of the chiral magnetic field $B_{5y}(x)$ obtained from the calculations (Bottom panel).
}
\label{fig:fig3}
\end{figure}

Here we take a magnetic domain wall configuration $\boldsymbol{m}(x)$ extended along the $x$-axis.
Owing to the strong perpendicular magnetic anisotropy in \ce{Co3Sn2S2},
the magnetization in each domain tends to point in the $\pm z$-direction.
The magnetic anisotropy field measured by experiments reaches up to $\mathrm{20\;\mathrm{T}}$~\cite{Liu2018,Shiogai2021},
which is also reproduced by our model calculation~(see Appendix \ref{sec:ma} for details of its calculation).
Therefore, here we consider a domain wall centered at $x=0$,
separating the two domains of $\boldsymbol{m} = \mp \hat{\boldsymbol{z}}$ for $x \gtrless 0$.
We take the N\'{e}el domain wall structure as schematically shown in Fig.~\ref{fig:fig3}(a),
\begin{align}
    \boldsymbol{m}(x) &= -\hat{\boldsymbol{z}} \tanh \frac{x}{w} + \hat{\boldsymbol{x}} \ \mathrm{sech} \frac{x}{w}, \label{eq:DW-Neel}
\end{align}
where $w$ is the domain wall width,
serving as $l_m$ in the condition of Eq.~(\ref{eq:local-uniform}).
Magnetization $\boldsymbol{m}$ inside this domain wall satisfies $\phi = 0^\circ$ with $\theta$ varied.
As seen in the previous section,
when the magnetization direction $\theta$ is varied with $\phi = 0^\circ$ fixed,
the Weyl point pair $\mathrm{WP}_1^\pm$ smoothly move in momentum space,
without forming a nodal ring.
%
Therefore, as long as the magnetization can be considered locally uniform for the Weyl fermions,
we may rely on the chiral gauge field picture.




By evaluating 
$\boldsymbol{K}_\eta$ for each position $x$,
and using the relations for $\boldsymbol{A}_5$ and $\boldsymbol{B}_5$ in Eqs.~(\ref{eq:A5-K}) and (\ref{eq:B5-A5}),
we calculate the spatial distribution of the chiral magnetic field $\boldsymbol{B}_5(x)$ around the domain wall for $\mathrm{WP}_1^\pm$.
Since $\mathrm{WP}_1^\pm$ are confined in $k_y = 0$ plane,
$\boldsymbol{A}_5(x)$ has only $x$ and $z$-components.
Therefore, in $\boldsymbol{B}_5(x) = (0, -\partial_x A_{5z}, \partial_x A_{5y}) $, only its $y$-component becomes nonzero,
as schematically shown in Fig.~\ref{fig:fig3}(a).

Figure \ref{fig:fig3}(b) shows the profile of $B_{5y}(x)$ obtained from our calculation,
which is localized around the domain wall center $x=0$.
We note that its profile is asymmetric, $B_{5y}(x) \neq B_{5y}(-x)$.
    This can be understood by comparing the Weyl point positions evaluated with ${\bm m}(x)$,
    which we denote as $\boldsymbol{K}_\eta |_{{\bm m}(x)}$,
    with those with ${\bm m}(-x)$, denoted as $\boldsymbol{K}_\eta |_{{\bm m}(-x)}$. 
The magnetization vectors 
$\boldsymbol{m}(\pm x) = \left(\mathrm{sech}\frac{x}{w}, 0, \mp \tanh\frac{x}{w}\right) $ satisfy the mapping
by the twofold rotation $R_{2x}$ around the $x$-axis,
$R_{2x}[\boldsymbol{m}(x)] = \boldsymbol{m}(-x)$.
However, since the \CSS crystal belongs to $R{\bar{3}}m$ and is not symmetric by $R_{2x}$,
the Weyl point positions do not satisfy the mapping by $R_{2x}$,
i.e., $R_{2x} [\boldsymbol{K}_\eta|_{\boldsymbol{m}(x)}] \neq \boldsymbol{K}_\eta|_{\boldsymbol{m}(-x)}$, which can be seen in Fig.~\ref{fig:fig1}(e).
In particular, for the $z$-component, it demands,
\begin{align}
    K_\eta^z |_{\boldsymbol{m}(x)} \neq -K_\eta^z |_{\boldsymbol{m}(-x)}.
\end{align}
    By applying Eqs.~(\ref{eq:A5-K}) and (\ref{eq:B5-A5}) to this inequality,
    we find that
asymmetry of $A_{5z}$ and $B_{5y} (= -\partial_x A_{5z})$,
\begin{align}
    A_{5z}(x) \neq -A_{5z}(-x), \quad
    B_{5y}(x) \neq B_{5y}(-x),
\end{align}
is permitted.



Let us focus on the magnitude of $\boldsymbol{B}_5$.
Since $\boldsymbol{A}_5$ here is the function of $x/w$ via $\boldsymbol{m}(x)$,
$B_{5y} = -\partial_x A_{5z}$ scales by $1/w$.
From our numerical calculations,
we see that its maximum magnitude scales as
\begin{align}
    |\boldsymbol{B}_5|_{\mathrm{max}} \approx \frac{2600}{w \mathrm{[nm]}} \;\mathrm{T}.
\end{align}
    Whereas there has been no direct measurement of the domain wall width $w$ in \CSS,
    here we set $w = 10\;\mathrm{nm}$ as its hypothetical length scale.
This scale is about one order of magnitude smaller than that in conventional ferromagnetic materials like permalloys $(w \sim 100\;\mathrm{nm})$,
    regarding the strong uniaxial anisotropy in \CSS.
With the assumed width $w = 10\;\mathrm{nm}$, 
the strength of $\boldsymbol{B}_5$ reaches
    \begin{align}
        |\boldsymbol{B}_5|_{\mathrm{max}} \approx 260 \;\mathrm{T}.
    \end{align}
This value is
about 10 times larger than that arising from lattice strain,
in a thin film of topological Dirac semimetal $\ce{Cd3As2}$ \cite{Liu2017PRB}.

Under such a large chiral magnetic field $|\boldsymbol{B}_5|$,
the Weyl electrons are confined within the scale of magnetic length
$l_\phi = (e|\boldsymbol{B}_5|)^{-1/2} \approx 1.6 \;\mathrm{nm}$,
which serves as $l_f$ in the condition of Eq.~(\ref{eq:local-uniform}).
Compared to $l_\phi$,
the variation of the magnetization is much moderate (i.e., $w \gg l_\phi$),
and hence the condition of Eq.~(\ref{eq:local-uniform}) is satisfied.
Therefore,
we can regard that the magnetization is locally uniform
so that we can rely on the chiral gauge field picture.
%

    The most important effect of the chiral magnetic field is
    the Landau quantization,
even without an ordinary magnetic field~\cite{Liu2013PRB,Grushin2016PRX,ilan2020pseudo}.
The energy-momentum dispersion of the zeroth Landau level (LL) is unidirectional along the chiral magnetic field ${\bm B}_5$
    for both chiralities,
    which are known as the chiral Landau levels.
        This is in a clear contrast to the Landau quantization by the ordinary magnetic field,
        where the zeroth LLs of the left- and right-handed fermions are dispersed oppositely to each other \cite{Yang2011prb}.
Under the ${\bm B}_5$, the level spacing between the zeroth and first LLs at the Weyl point becomes
$\delta E = v_{\rm F} \sqrt{2e|\boldsymbol{B}_5|}$,
where $v_{\rm F}$ is the Fermi velocity around the Weyl point.
In \CSS, $v_{\rm F}$ = 0.5$~$eV\AA $~$is reported by the photoemission spectroscopy measurements \cite{Liu2018,Xu2018},
from which we obtain the level spacing $\delta E \approx 45 \;\mathrm{meV}$ under $|{\bm B}_{5}| \approx 260\;\mathrm{T}$.
This value is comparable to the Fermi energy $E_{\rm F}=60 \; \mathrm{meV}$ with respect to the Weyl points in \CSS~\cite{Liu2018,Liu2019}.
Therefore, $E_{\rm F}$ crosses only a few LLs including the zeroth LLs,
and we can regard that the electron system is in the quantum regime.
We expect that the zeroth chiral LLs give the dominant contribution to the anomalous magnetoelectric responses
around the domain wall,
as we shall discuss in the next section.

\section{Magnetoelectric effect at domain wall} \label{sec:discussion}

So far we have observed the structure of the chiral magnetic field around the domain wall.
In this section, we discuss the magnetoelectric effect that arises from the chiral electromagnetic fields.
%
We consider the charge pumping induced by the domain wall motion,
which was first predicted in spin-momentum locked Weyl semimetals
based on the idea of the chiral electromagnetic fields \cite{Araki2018prapplied,Araki2020adp}.

Before proceeding to the detailed discussion for \CSS,
we first review the general concept of the charge pumping effect in magnetic Weyl semimetals.
When one applies a magnetic field or injects an electric current to the system,
the magnetic moments forming the domain wall collectively get rotated time-dependently.
As a result, the position of the domain wall gets eventually shifted.
In this process,
since the magnetization $\boldsymbol{m}(\boldsymbol{r},t)$ is time-dependent, 
the chiral gauge field ${\bm A}_5(\boldsymbol{r},t)$ also becomes time-dependent.
It thus yields the chiral electric field $\boldsymbol{E}_5 = -\partial_t \boldsymbol{A}_5$
localized at the domain wall.
Suppose that the domain wall is sliding along the $x$-axis with the velocity $V_{\mathrm{DW}}$ without changing its internal structure.
In this case, the magnetization and the chiral gauge field can be written as
$\boldsymbol{m}(\boldsymbol{r},t) = \boldsymbol{m}(x')$ and
$\boldsymbol{A}_5(\boldsymbol{r},t) = \boldsymbol{A}_5(x')$,
    with $x' = x-V_{\mathrm{DW}}t$.
From this $\boldsymbol{A}_5$, both the chiral electric and magnetic fields emerge,
\begin{align}
    \boldsymbol{E}_5(x') &= -\partial_t \boldsymbol{A}_5(x') = V_{\mathrm{DW}} \partial_{x'} \boldsymbol{A}_5(x'),
    \label{eq:E5A5} \\
    \boldsymbol{B}_5(x') &= \boldsymbol{\nabla} \times \boldsymbol{A}_5(x') = \hat{\boldsymbol{x}} \times \partial_{x'} \boldsymbol{A}_5(x').
    \label{eq:B5A5}
\end{align}
We can see that $\boldsymbol{E}_5$ is perpendicular to the chiral magnetic field $\boldsymbol{B}_5$, satisfying the relation,
\begin{align}
    \boldsymbol{B}_5 = \hat{\boldsymbol{x}} \times \frac{\boldsymbol{E}_5}{V_{\mathrm{DW}}}. \label{eq:B5-E5}
\end{align}

As a result,
similarly to the conventional electromagnetic fields,
the chiral electromagnetic fields give rise to the ordinary Hall current \cite{Araki2018prapplied},
\begin{align}
    \boldsymbol{j}^{(H)} &= \sigma_H \frac{\boldsymbol{B}_5}{|\boldsymbol{B}_5|} \times \boldsymbol{E}_5.
     \label{eq:JH}
\end{align}
Here $\sigma_H$ is the Hall conductivity summed over the two chiralities.
The Hall current is localized at the domain wall, and flows along the $x$-axis in parallel with the domain wall motion.
We should note that this current is independent of the polarity of the domain wall,
because both $\boldsymbol{B}_5$ and $\boldsymbol{E}_5$ flip their signs
when the magnetization direction $\boldsymbol{m}(\boldsymbol{r},t)$ is reversed.
    In contrast to the adiabatic pumping by the Weyl point motion allowed in noncentrosymmetric Weyl semimetals \cite{Harada2023prb,Ishizuka2016prl,Ishizuka2017prb},
    the present effect is allowed even though the crystal structure is centrosymmetric, like \CSS.

We now apply the above discussions to the domain wall in \CSS,
with the structure of the chiral magnetic field calculated in the previous section.
Since the above theory requires only the presence of ${\bm B}_5$ and ${\bm E}_5$,
it can be applied to the Weyl fermions without spin-momentum locking.
To realize the domain wall motion,
one needs to switch the magnetization around the domain wall,
by applying an external magnetic field or a current-induced spin torque to exceed the coercive field.
At $T = 100 \;\mathrm{K}$ in ferromagnetic \CSS,
the coercive field is suppressed below $50 \;\mathrm{mT}$ in the bulk sample \cite{Liu2018},
whereas it reaches up to $5\;\mathrm{T}$ in the film sample of the thickness $40\;\mathrm{nm}$ \cite{Shiogai2021}.
Therefore, the domain wall motion is more likely to occur in the bulk \CSS,
while it is strongly suppressed in thin films.

For the domain wall moving in the $x$-direction,
$\boldsymbol{E}_5$ and $\boldsymbol{B}_5$ mainly point to the $z$- and $y$-directions, respectively,
because $\boldsymbol{A}_5$ is dominated by the $z$-component of the magnetization for the N\'{e}el domain wall in \CSS.
If the domain wall of the width $w = 10 \;\mathrm{nm}$ is sliding with the velocity $V_{\mathrm{DW}} = 100 \;\mathrm{m/s}$,
we obtain a fairly strong chiral electric field,
\begin{align}
    |E_{5z}| = V_{\mathrm{DW}} |B_{5y}| \approx 26 \; \mathrm{kV/m},
\end{align}
from the relation Eq.~(\ref{eq:B5-E5}).
Since the electron system here is in the quantum regime under the strong $\boldsymbol{B}_5$,
the current $\boldsymbol{j}^{({\rm H})}$ is regarded as the quantum Hall current from the zeroth LLs,
which is free from dissipation by disorder scattering.
The quantum Hall conductivity for the Weyl fermions becomes \cite{Yang2011prb}
\begin{align}
    \sigma_H = N_{\rm W} \frac{E_{\rm F}}{v_{\rm F}} \frac{e^2}{4\pi^2},
\end{align}
where $N_W$ accounts for the number of the Weyl nodes around the Fermi level ($N_{\rm W} =6$ in \CSS).
By using the values of $|E_{5z}|$ and $\sigma_H$ given above,
the 
Hall current
can be estimated from Eq.~(\ref{eq:JH}) as
\begin{align}
    j^{({\rm H})}_x \approx -1.2 \;\mathrm{nA/nm^2},
\end{align}
which is present inside the domain wall.
Therefore, the Hall voltage arising between the two sides of a single domain wall becomes
\begin{align}
    V_x = \int_{\mathrm{DW}} dx \ \rho_{xx} j^{({\rm H})}_x \approx \rho_{xx} j^{({\rm H})}_x w = 55 \;\mathrm{\mu V},
\end{align}
where we have used the longitudinal resistivity $\rho_{xx} = 2.9 \times 10^{-4} \; \mathrm{\Omega} \; \mathrm{cm}$
measured in bulk \CSS \cite{Ikeda2021}.
    Since this Hall voltage is independent of the domain wall polarity,
    the voltage in a multi-domain system is multiplied by the number of domain walls.

    The mechanism of charge pumping discussed above is distinct from the so-called spinmotive force (or emergent electric field),
    which is the voltage induced by magnetization dynamics in metals with nontopological electrons \cite{Volovik1987jopc,Barnes2007prl,nagaosa2012physscr}.
    The spinmotive force arises from the spin Berry phase,
    which is adiabatically accumulated by the electron spin following the magnetization dynamics.
    On the other hand, the charge pumping of Weyl electrons arises from the drastic changes of the band structure including the Weyl points.
    There is also a quantitative difference between these two mechanisms.
    In conventional ferromagnets,
    the spinmotive force from the sliding motion of domain wall, with $w=10~{\rm nm}$ and $V_{\rm DW}=100~{\rm m/s}$,
    scales $V_{\mathrm{SMF}} \approx 66 \;\mathrm{nV}$ (see Appendix \ref{sec:smf} for its derivation)
    \cite{Saslow2007prb,Duine2008prb,Duine2009prb,Tserkovnyak2008prb}.
    In comparison with this spinmotive force contribution,
    the Hall voltage from the chiral electromagnetic fields discussed here is about 800 times larger.

\section{Conclusion}


In this article,
we have studied
    the structure of the chiral gauge field for the Weyl fermions in \CSS.
    Its electronic structure is fully spin polarized and does not have the spin-momentum locking structure.
    By using the tight-binding model for the low-energy band structure,
    we have investigated the positional shift of the Weyl points depending on the magnetization direction.
Since the shift of the Weyl points serves as the chiral gauge field for the Weyl fermions,
we have demonstrated the strong chiral magnetic field appearing around a magnetic domain wall.
We find that the chiral magnetic field reaches up to $260 \;\mathrm{T}$ in \CSS,
which is because the Weyl points positions are highly sensitive to the magnetization direction.
With such a giant chiral magnetic field arising at the domain wall,
there appear the chiral Landau levels that are localized at the domain wall.
    These chiral Landau levels reflect the topological nature of the Weyl fermions,
    and are capable of yielding the anomalous magnetoelectric responses
    beyond the Maxwell's equations.

To study the outcome of the chiral gauge field picture associated with the magnetic texture,
we have considered the sliding motion of the domain wall.
Since the sliding domain wall accompanies both chiral magnetic and chiral electric fields,
it induces a Hall current and voltage along the sliding direction.
The magnitude of the Hall voltage scales
about 800 times larger than that from the conventional spinmotive force in nontopological magnets.
Our finding offers one concrete example of the interplay between magnetism and electron transport
in a realistic Weyl semimetal material, 
even though the spin-momentum locking is absent.
Such an effect would be helpful in detecting the magnetic textures as carriers of information,
in future spintronics devices such as memories.
To enable the power-efficient operation of such spintronics devices,
one also needs to manipulate the magnetic textures electrically by spin torque~\cite{kurebayashi2019theory,kurebayashi2021jpsj,Araki2021prl,yamanouchi2022sciadv,wang2023magnetism}.
The theoretical understanding of its process in \CSS is left for future studies.

\acknowledgments
The authors would like to appreciate 
K.~Fujiwara,
J.~Ieda,
Y.~Kato,
K.~Kobayashi,
Y.~Motome,
K.~Nakazawa,
T.~Oka, 
and
A.~Tsukazaki,
for valuable discussions.
This work was supported by
JST CREST, Grant No.~JPMJCR18T2.
Y.~A. was supported by 
JSPS KAKENHI, Grant No.~22K03538. 
A.~O.~was supported by
JST CREST, Grant No.~JPMJCR19T3 
and by JSPS KAKENHI. Grant No.~JP23K19194 

\appendix

\section{Chiral gauge field in Weyl semimetals} \label{sec:chiral-gauge}
    Here we review the fundamental idea of the chiral gauge field based on Refs~\cite{Liu2013PRB,ilan2020pseudo}.
    Around each Weyl point with the index $\eta$,
    the low-energy Hamiltonian is symbolically written as
    \begin{align}
        H_\eta(\boldsymbol{k}) &= E_\eta + \sum_{\mu = x,y,z} v_{\eta,\mu} (\boldsymbol{k} - \boldsymbol{K}_\eta)_{\mu} \tau_{\eta,\mu} .
    \end{align}
    Here $E_\eta$ is the energy level of the Weyl point,
    $\boldsymbol{v}_\eta$ is the group velocity around the Weyl point,
    and $\boldsymbol{K}_\eta$ is the position of the Weyl point in momentum space.
    The Pauli matrix $\boldsymbol{\tau}_\eta$ accounts for the internal degrees of freedom (i.e., spin, orbital, sublattice, etc.)
    that correspond to the two bands forming the Weyl point.
    Since $\boldsymbol{K}_\eta$ appears as the shift in momentum $\boldsymbol{k}$,
    its effect can be regarded as the effective vector potential
    $\boldsymbol{A}_\eta = \frac{1}{e} \boldsymbol{K}_\eta$ for the low-energy excitations around the Weyl point.

    Here we focus on a pair of Weyl points with chiralities $\eta=\pm$.
    The chirality-odd part of $\boldsymbol{A}_\pm$,
    \begin{align}
        \boldsymbol{A}_5 = \frac{1}{2} \left( \boldsymbol{A}_+ - \boldsymbol{A}_- \right),
    \end{align}
    is called the chiral (axial) gauge field, in the context of the relativistic quantum mechanics.
    The chiral gauge field $\boldsymbol{A}_5$ is associated with the U(1) chiral gauge transformation
    $U(\phi) = e^{\pm i\phi}$,
    which rotates the phases of the Weyl fermions of the chiralities $\pm$ by the opposite angles $\pm \phi$.
    If the system possesses spatial inversion symmetry, like in ferromagnetic \CSS,
    $\boldsymbol{K}_+$ and $\boldsymbol{K}_-$ are related by inversion operation.
    i.e., $\boldsymbol{K}_+ \equiv - \boldsymbol{K}_-$ modulo the reciprocal lattice vectors.
    Therefore, we focus on the chirality-odd part $\pm \boldsymbol{A}_5$ in $\boldsymbol{A}_\pm$.

    The Weyl point positions $\boldsymbol{K}_\pm$ generally depend on the material parameters,
    such as lattice strain, chemical substitution rates, or magnetic orderings.
    Therefore, the pertrubation of those parameters leads to the shift in $\boldsymbol{K}_\pm$,
    and effectively generates $\boldsymbol{A}_5$ for the Weyl fermions.
    As in the case of the ordinary U(1) gauge field $\boldsymbol{A}$ for electromagnetism,
    the constant chiral gauge field $\boldsymbol{A}_5$ has no physical effect on the Weyl fermions.
    When the perturbation that yields $\boldsymbol{A}_5$ is temporally or spatially modulated, 
    $\boldsymbol{A}_5$ comes into effect.
    If $\boldsymbol{A}_5$ has a time-dependent modulation, it yields the electric component,
    \begin{align}
        \boldsymbol{E}_5 = -\partial_t \boldsymbol{A}_5(t),
    \end{align}
    which is called the \textit{chiral (axial) electric field}.
    If $\boldsymbol{A}_5$ has a spatial modulation, it yields the magnetic component,
    \begin{align}
        \boldsymbol{B}_5 = \boldsymbol{\nabla}\times \boldsymbol{A}_5(\boldsymbol{r}),
    \end{align}
    which is called the \textit{chiral (axial) magnetic field}.
    The chiral electromagnetic fields $(\boldsymbol{E}_5,\boldsymbol{B}_5)$ act on the Weyl fermions
    just like the ordinary electromagnetic fields $(\boldsymbol{E},\boldsymbol{B})$,
    but with the signs $\pm$ depending on the fermion chirality.

\section{Calculation of magnetic anisotropy} \label{sec:ma}

    In this section, we explain our calculation process of the magnetic anisotropy from the tight-binding model,
    which is briefly shown in Sec.~\ref{sec:weyl-trajectory}.
    This scheme was employed also in our previous literature \cite{Ozawa2019,Ozawa2022}.

    The magnetic anisotropy energy for the uniaxial anisotropy is phenomenologically written as
    \begin{align}
        E_{\mathrm{ani}}[\{\boldsymbol{m}_i\}] &= -\frac{1}{2}K \sum_{i} (m_i^z)^2,
    \end{align}
    which is summed over the magnetic moments $\{ \boldsymbol{m}_i \}$ in the system.
    The sign of the coefficient $K$ is positive for the easy-axis anisotropy.
    We here quantify the anisotropy in terms of the effective magnetic field,
    \begin{align}
        B_{\mathrm{ani}}^z &=-\frac{\partial E_{\mathrm{ani}}}{\partial m_i^z} = K m_i^z,
    \end{align}
    which is given in the unit of Tesla.

\begin{figure}[t]
    \centering
    \includegraphics[width=1.0\hsize]{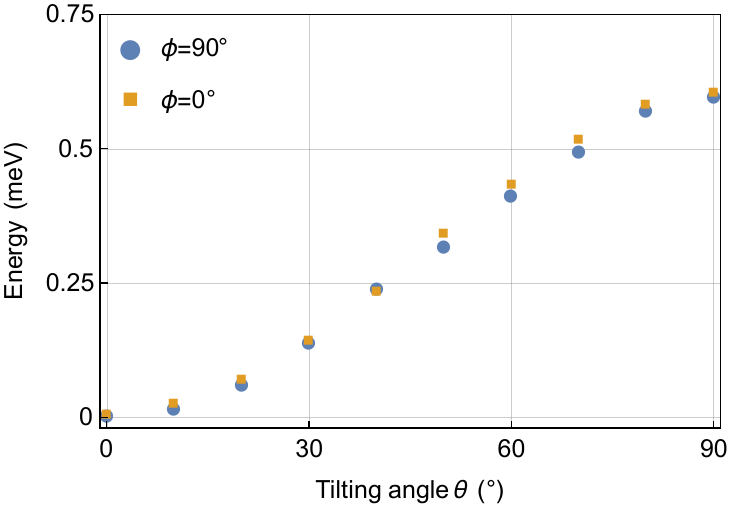}
    \caption{
    Magnetic anisotropy energy $E_{\mathrm{ani}}$ per a single primitive unit cell,
    computed from the tight-binding model.
    The tilting angle $\theta$ is varied, for the two different azimuthal angles $\phi=0^{\circ}$~(orange) and $\phi=90^{\circ}$~(blue).
    }
    \label{fig:figap1}
\end{figure}

    From the tight-binding model for the electron system,
    the anisotropy energy is given as the energy of electrons summed over the occupied states,
    \begin{align}
        E_{\mathrm{ani}}[\{\boldsymbol{m}_i\}] = \frac{1}{N}\sum_{n {\bm k}} f(E_{n {\bm k}}-E_F) E_{n {\bm k}},
    \end{align}
    under the configuration of magnetic moments $\{\boldsymbol{m}_i\}$.
    The momentum summation runs over the first Brillouin zone,
    with $N$ the number of meshes taken in the Brillouin zone,
    so that $E_{\mathrm{ani}}$ accounts for the anisotropy energy per a single (primitive) unit cell.
    We here calculate $E_{\mathrm{ani}}$ under the uniform ferromagnetic configurations $\boldsymbol{m}_i = \boldsymbol{m}$.
    We fix the electron number to charge neutrality~\cite{Ozawa2019},
    and hence the value of $E_F$ depends on the magnetization direction $\boldsymbol{m}$.

    By varying the polar angle $\theta$ of $\boldsymbol{m}$,
    we obtain the anglular dependence of $E_{\mathrm{ani}}$ shown in Fig.~\ref{fig:figap1}.
    We calculate $E_{\rm ani}$ for different azimuthal angles $\phi=0^\circ$ and $\phi=90^\circ$.
    We find the difference between them is negligible. 
    Comparing $E_{\mathrm{ani}}$ for $\boldsymbol{m}\parallel z \ (\theta = 0^\circ)$ and $\boldsymbol{m}\parallel x \ (\theta = 90^\circ)$,
    we obtain
    \begin{align}
        E_{\mathrm{ani}}[\boldsymbol{m}\parallel x] - E_{\mathrm{ani}}[\boldsymbol{m}\parallel z] = \frac{3}{2}K |\boldsymbol{m}|^2 = 0.6 \;\mathrm{meV},
    \end{align}
    where the prefactor $3$ accounts for the number of Co sites in the unit cell.
    By using $|\boldsymbol{m}| = 0.3 \mu_B$ for each Co site in \CSS,
    we obtain
    \begin{align}
        B_{\mathrm{ani}}^z &= K|\boldsymbol{m}| = 20 \;\mathrm{T}.
    \end{align}

\section{Spinmotive force in conventional ferromagnet} \label{sec:smf}
In this section, we briefly discuss how the domain wall motion in conventional ferromagnetic metals induces a voltage,
by the idea of the spinmotive force~\cite{Volovik1987jopc,Barnes2007prl}.
We also estimate the magnitude of the induced voltage in conventional ferromagnetic metals.

We consider a ferromagnetic metal in which the electron spins are polarized with the ratio $p (\leq 1)$.
If there is a dynamics in a magnetic texture $\boldsymbol{m}(x,t)$,
a conduction electron feels the effective electric field,
\begin{align}
    E_x^\pm = \pm \frac{p}{2e}\left[ \boldsymbol{m}\times \partial_t \boldsymbol{m} + \beta \partial_t \boldsymbol{m} \right]\cdot \partial_x \boldsymbol{m},
\end{align}
which is known as the spinmotive force.
Here $\pm$ denotes whether the electron belongs to the majority or minority spin state.
The first term comes from the Berry phase accumulated by the time evolution of the electron state,
when the electron spin adiabatically follows the magnetization dynamics \cite{Volovik1987jopc,Stern1992prl,Barnes2007prl}.
The second term denotes the nonadiabatic contribution,
which comes from the spin-flip process between the majority and minority spin states \cite{Saslow2007prb,Duine2008prb,Duine2009prb,Tserkovnyak2008prb}.
The parameter $\beta$ characterizes the nonadiabaticity,
which also appears in the formula for spin-transfer torque \cite{Zhang2005prl,Barnes2005prl,thiaville2005micromagnetic}.
From theoretical calculations,
the value of $\beta$ is estimated to be of the same order as the Gilbert damping parameter $\alpha$ for the magnetization dynamics \cite{Tserkovnyak2006prb,Kohno2006jpsj}.
Both $\alpha$ and $\beta$ scale around 0.01 in conventional ferromagnetic metals.

Let us consider the sliding motion of a Bloch domain wall,
$\boldsymbol{m}(x,t) = \left( 0, \ \mathrm{sech} \tfrac{x-V_{\mathrm{DW}}t}{w}, \ \tanh \tfrac{x-V_{\mathrm{DW}}t}{w} \right)$.
The adiabatic term vanishes for the sliding motion,
and we are left with the nonadiabatic term,
\begin{align}
    E_x^+(x,t) &= -\frac{\beta p}{2e} \frac{V_{\mathrm{DW}}}{w^2} \mathrm{sech}^2 \; \frac{x-V_{\mathrm{DW}}t}{w},
\end{align}
for the majority spin.
Therefore, the voltage arising between the two sides of a single domain wall is given as
\begin{align}
    V_x^+ = -\int dx \ E_x^+(x,t) = \frac{\beta p}{e} \frac{V_{\mathrm{DW}}}{w}.
\end{align}
By taking the nonadiabaticity parameter $\beta \approx 0.01$,
and assuming the full spin polarization $(p=1)$,
    we obtain $|V_x^+| \approx 66 \;\mathrm{nV}$ for a single domain wall
    with $w = 10 \;\mathrm{nm}$ and $V_{\mathrm{DW}} = 100 \;\mathrm{m/s}$.

\bibliography{ref}
\end{document}
